\documentclass[superscriptaddress,twocolumn,showpacs,amsmath,amssymb,floats,pre]{revtex4}

\usepackage{graphicx}

\usepackage{dcolumn}

\usepackage{bm}

\usepackage{revsymb}

\usepackage[usenames]{color}

\usepackage{subfigure}

\usepackage{bigints}

\newcommand{\Xv}{\bm{{\rm X}}}

\newcommand{\rv}{\bm{{\rm r}}}

\newcommand{\as}[1]{{#1}}

\begin{document}

\title{Beads on a String: Structure of bound aggregates of globular particles and long polymer chains}
\author{Anton Souslov}
\affiliation{School of Physics,
Georgia Institute of Technology, Atlanta, GA, 30332, USA }
\author{Jennifer E.~Curtis}
\affiliation{School of Physics,
Georgia Institute of Technology, Atlanta, GA, 30332, USA }
\affiliation{Parker H.~Petit Institute for Bioengineering and Bioscience,
Georgia Institute of Technology, Atlanta, GA, 30332, USA }
\author{Paul M.~Goldbart}
\affiliation{School of Physics,
Georgia Institute of Technology, Atlanta, GA, 30332, USA }

\date{\today}

\begin{abstract}
Macroscopic properties of suspensions, such as those composed of globular particles (\textit{e.g.}, colloidal or macromolecular), can be tuned by controlling the equilibrium aggregation of the particles. We examine how aggregation -- and, hence, macroscopic properties -- can be controlled in a system composed of both globular particles and long, flexible polymer chains that reversibly bind to one another. We base this on a minimal statistical mechanical model of a single aggregate in which the polymer chain is treated either as ideal or self-avoiding, and, in addition, the globular particles are taken to interact with one another via excluded volume repulsion. Furthermore, each of the globular particles is taken to have one single site to which at most one polymer segment may bind. Within the context of this model, we examine the statistics of the equilibrium size of an aggregate and, thence, the structure of dilute and semidilute suspensions of these aggregates. We apply the model to biologically relevant aggregates, specifically those composed of macromolecular proteoglycan globules and long hyaluronan polymer chains. These aggregates are especially relevant to the materials properties of cartilage and the structure-function properties of perineuronal nets in brain tissue, as well as the pericellular coats of mammalian cells.
\end{abstract}


\maketitle

\section{Introduction}
For suspensions composed of long polymer chains and colloidal particles, a wealth of materials properties
can be explored by tuning the polymer-colloid interactions.
For example, in suspensions in which the polymers and the colloids do not bind to one another, flocculation of large aggregates can 
be controlled by tuning the polymer concentration, which determines the effective depletion or electrostatic forces between colloidal particles 
(as reviewed in, \textit{e.g.}, Refs.~\cite{Jenkins1996,Tuinier2003}).
In other systems, polymer-colloid interactions lead to the formation of polymer bridges between colloidal particles, 
which results in macroscopic gelation or flocculation (as reviewed in, \textit{e.g.}, Ref.~\cite{Dickinson1991}).
A similar bridging mechanism leads to gelation in polymer-micelle systems~\cite{Cabane1977,Goddard1986,Hansson1996}.
Such bridging phenomena have been modeled in terms of the adsorption 
of polymer chains onto the surfaces of suspended particles~\cite{Alexander1977,Pincus1984}.
In this work, we instead consider a suspension in which polymer bridges form between globular suspended particles 
due to the presence on each such particle of a \emph{single specific site}
to which the polymer binds. For this system, aggregation is limited by the polymer chain length
and, therefore, the aggregates are limited to the microscale and, thus, do not flocculate.
We consider the case for which the polymer-colloid binding is reversible. 
For this case, we describe how it may be possible to control experimentally the characteristic size of aggregates, 
for example by tuning the length of the polymer chains, the temperature, or the density of colloidal particles.
The size of aggregates then determines the macroscopic materials properties of the suspension.

To develop a statistical mechanical approach to the structure of these aggregates, 
we model the colloidal particles as hard spheres, and the polymers as Gaussian chains, possibly with self-avoidance.
By considering key physical features of polymer-colloid suspensions -- the steric repulsion between the colloidal particles
and the presence on each particle of exactly one binding site for polymer chains -- we 
derive results for the statistics of the number $M$ of colloidal particles attached to each polymer chain and, consequently,
of the linear size $R$ of the aggregate.
Thus, for self-avoiding polymer chains we derive the following expressions for the mean and variance of $M$:
\begin{align}
\langle M \rangle \sim & \left(\beta (\mu+E_b) \right)^{2/3} \left( R_g/2 a \right)^{5/3} \nonumber \\
& \,\,\,\,\,\, = \left(\beta (\mu+E_b)\right)^{2/3} N \left( \ell_0/2 a\right)^{5/3}, \\
\langle (\delta M)^2 \rangle \sim & \left(\beta (\mu+E_b)\right)^{-1/3} \left( R_g/2 a \right)^{5/3},
\label{eq:res}
\end{align}
in terms of $\beta = 1/ k_B T$ (where $T$ is the temperature), the chemical potential $\mu$ of the colloidal particles,
the polymer-colloid binding energy $E_b$, the characteristics of an isolated polymer chain (\textit{i.e.}, the segment length $\ell_0$, the number of segments $N$, and the radius of gyration $R_g = N^{3/5} \ell_0$), and the radius $a$ of a colloidal particle.
The relative width of the distribution [which is given by $\langle \left(\delta M\right)^2 \rangle /\langle M \rangle^2 \sim \left(\beta (\mu+E_b)\right)^{-5/3}$]
decreases as the binding energy (and, thus, the size of the aggregate) increases. 
As a consequence, for sufficiently large aggregates
the size distribution is sharply peaked around the mean, and the aggregates may be addressed within one of the two essentially equivalent
ensembles -- canonical at fixed $M$, or grand canonical at fixed $\mu$.
In suspension, such aggregates may themselves be modeled in terms of effective self-avoiding chains, each composed of $M$ segments of segment length $2a$, so that the typical linear size $R$ of an aggregate is given by $a M^{3/5}$. 

The physical system that inspired this model is the suspension of biomolecular aggregates composed of long chains of the polysaccharide hyaluronan (HA, which behaves as a polymer chain~\cite{Cowman2005}) and the types of globular macromolecules called proteoglycans 
(PG, \textit{e.g.}, aggrecan, versican, brevican and neurocan -- collectively known as lecticans~\cite{Wight2011}) that bind to HA. 
\as{Although the ``beads on a string'' language has been used to describe the euchromatin structure of nucleosome beads along a string of DNA, we do not intend to describe that system in this work.}
These aggregates are found in numerous biological contexts, including the extracellular matrix (ECM)~\cite{Toole2004, Yamaguchi2000} and the cell-surface-associated polymer matrix called the pericellular coat~\cite{Boehm2009, McLane2013}.  
For example, in cartilage, a high concentration of aggrecan (which is bound to hyaluronan strands and therefore aggregated) works in concert with a dense network of collagen fibrils to form a tough and resilient load-bearing material. The aggrecan attracts water into the tissue by osmosis, thereby exerting a swelling pressure on the collagen network and enabling the cartilage to resist large compressive loads (see \textit{e.g.}, Ref.~\cite{Hardingham1998}).
As another example, when bound to the surfaces of cells, PG aggregates co-regulate cell behavior, especially processes that involve changes in cell adhesion, such as migration~\cite{Ricciardelli2007}, proliferation~\cite{Evanko1999}, and synaptogenesis~\cite{Wang2012}.

\begin{figure*}[thp]
\includegraphics[angle=0]{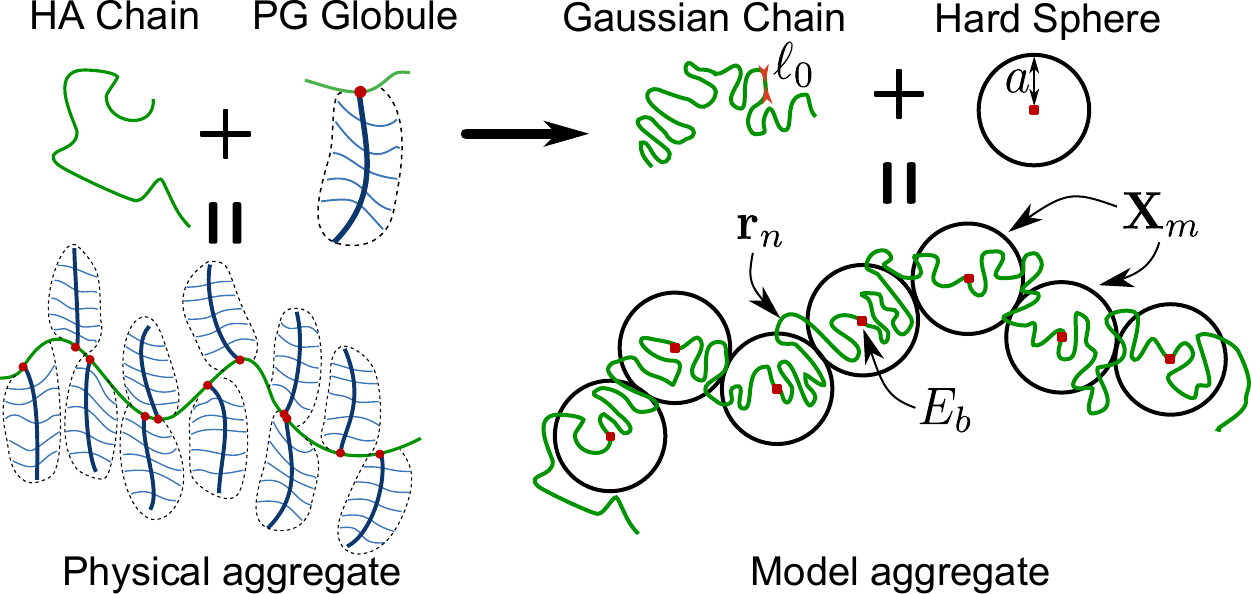}
  \caption{A schematic diagram of the model presented in this work and its relation
to molecular aggregates composed of hyaluronan (HA) and proteoglycans (PG).
The variables $\rv_n$ and $\Xv_m$ correspond
to the positions of the chain segments and the globule centers,
respectively. The model parameters include $E_b$ -- the binding energy,
$\ell_0$ -- the chain segment length, and $a$ -- the globule size.  
The PG globules, which are bottle-brush polymers that avoid overlap,
are modeled as hard spheres and the polysaccharide HA chain is modeled
as a Gaussian chain.
These two components aggregate due to the affinity of the globules to bind to the chain.
}
\label{fig:model}
\end{figure*}

In this biological context, the parameters in Eq.~(\ref{eq:res}) take on specific values or ranges of values:
\begin{itemize}
\item For HA chains, we choose the segment length to be twice the persistence length, \textit{i.e.}, $\ell_0 \sim 14\,{\rm nm}$~\cite{Cowman2005},
and the contour length $N \ell_0$ to be up to $25\,\mu{\rm m}$~\cite{Toole2004},
such that $N \alt 10^3$ segments.
\item We choose $a \sim 100\,{\rm nm}$ to correspond to the size of the PG molecules that attach to the HA chain. 
In real systems, the PG molecules are somewhat anisotropic bottlebrushes, having cylindrical dimensions 
${\sim 100\,{\rm nm}\times100\,{\rm nm}\times350\,{\rm nm}}$ when fully stretched~\cite{Ng2003,Ng2004}
but, for simplicity, we do not include this anisotropy in the model. 
We expect that, as a consequence of thermal fluctuations, the typical shape of the PG molecules in solution is less anisotropic.
In order to model globular bottle-brush polymers that are highly swollen as a result of electrostatic repulsion between their highly charged side-chains (see Fig.~\ref{fig:model})~\cite{Hardingham1974, Nap2008}, we include hard-core repulsion between the globules. In biological contexts, the Debye length is sufficiently short ($\sim1\,{\rm nm}$)
 that the only relevant interaction is the strong, steric repulsion that suppresses PG-PG overlap.
\item The binding site of each PG molecule occupies ${\sim10\,{\rm nm}}$ of HA contour length, \textit{i.e.}, a length \mbox{$\sim\ell_0$}~\cite{Morgelin1988}.
\item PG-HA aggregates in cartilage are composed of up to $\sim 100$ PG globules bound to a single HA chain~\cite{Morgelin1988}.
\end{itemize}

\label{sec:intro}

\section{Model and Results} 
\label{sec:model}
We begin by giving a detailed account of the models we shall use to describe a composite system consisting of a single polymer chain and the globular particles that bind to it.
Next, we construct the partition function for a single aggregate. 
We then simplify this partition function to obtain our main result: the distribution of aggregate sizes,
by which we mean the distribution of $M$ as well as of $R$.

We emphasize the interplay between chain entropy and the binding statistics of globules,
and also emphasize associated phenomena occurring on lengthscales that are much larger than the HA persistence length.
At these large lengthscales, a chain model is sufficient to capture the physical properties of the polymer,
and we consider both the models of Gaussian and self-avoiding chains.
For a Gaussian chain, the polymer segments are modeled by Hookean springs (of natural length $\ell_0$), which connect 
universal joints at positions $\rv_n$ (with $n = 1,\ldots, N^\prime$).
Self-avoiding chains, in addition, self-interact, via the potential $U_{n,n^\prime} \equiv U(|\rv_n - \rv_{n^\prime}|)$,
which is relevant for physical polymers in a good solvent.
Thus, the effective free energy of chain configuration $\{\rv_n\}$ is given by
\begin{equation}
G_{N^\prime}(\{\rv_n\}) = k_B T \sum_{n} \frac{|\rv_n - \rv_{n-1}|^2}{2 \ell_0^2} + \sum_{n < n^\prime}U_{n,n^\prime}.
\nonumber
\end{equation}
We use this chain model only for sufficiently small stretching of the chain. 
The opposite case of chains so taut as to resist further stretching through
molecular binding energy, rather than through entropic elasticity,
requires more detailed, chain-specific, considerations.

We model the globular proteoglycans as particles at positions $\Xv_m$ (with $m = 1,\ldots, M$), interacting
with one another via a two-body potential.
We assume the effects of the inherent anisotropy of these molecules to be negligible, 
and thus model their interactions via a central-force potential $V$.
Furthermore, we approximate $V$ as a short-range, hard-sphere potential of radius $a$, 
to account for the mutual impenetrability of the globules, which in essence holds at the relevant energy scales:
\begin{equation}
\nonumber
V_{m,m^\prime} \! \equiv \! V_{\mathrm{hs}}(|\Xv_m - \Xv_{m^\prime}|) \! = \!\left\{
     \begin{array} {cl}
       0, & \mathrm{for\,\,} |\Xv_m - \Xv_{m^\prime}| > 2 a; \\
       \infty, &  \mathrm{otherwise}.
     \end{array} 
     \right.
\end{equation}

The partition function for one aggregate factors in all configurations of the chain segments $\{\rv_n\}$ and of the hard spheres $\{\Xv_m\}$,
as well as the locations $\{\sigma_m\}$ of the binding sites of the $m^\mathrm{th}$ hard sphere along the chain backbone 
(which we may take to be ordered, \textit{i.e.}, $\sigma_m < \sigma_{m+1}$ for all $m$).
Each $\sigma_m$ may take on values in $\{ 1, \ldots, n \}$, and is associated with the binding constraint $\Xv_m = \rv_{\sigma_m}$.
Thus, in terms of the chemical potential $\mu$ of the suspended globules and the binding energy $E_b$, the Boltzmann factor for creating a globule bound to a monomer of the chain is $e^{\beta (\mu + E_b)}$~\as{\cite{f2}}.
For an aggregate formed in a sufficiently dilute suspension of globules, $\mu$
is related to the density of unbound globules $\rho_u$ via the expression $\mu = k_B T \ln \left( \rho_u a^3 \right)$, 
obtained from the analogous expression for the chemical potential of an ideal gas. 
In this (grand canonical) ensemble (with respect to the globules), the partition function of an aggregate may be expressed as 
\begin{equation}
\label{eq:Z}
Z = \sum_M e^{\beta (\mu + E_b) M} Z_M,
\end{equation}
where $Z_M$ is the partition function for an aggregate with a fixed number $M$ of bound globules (\textit{i.e.}, in the canonical ensemble) and is given by
\begin{widetext}
\begin{equation}
\label{eq:zm}
Z_M = {\sum_{\{\sigma_m\}}}^{\!\!\prime} \bigintsss \left[\prod_{n=1}^{N^\prime} d \rv_n\right] \left[\prod_{m = 1}^M d \Xv_m\right] \left( \prod_{m < m^\prime} e^{-\beta V_{m,m^\prime}} \right) 
e^{- \beta G(\{\rv_n\})}
\prod_{m = 1}^M \delta^{(3)}\left(\Xv_m - \rv_{\sigma_m}\right),
\end{equation}
\end{widetext}
in which $\sum^\prime$ is the sum over all ordered binding locations ${\{\sigma_m\}}$.
Thus, this partition function considers all of the configurations of a single aggregate, and assigns each configuration the appropriate weight.
We use the assumptions that $N^\prime \gg 1$ and $M \gg 1$, so we need not consider polymer end effects. 
Thus, we replace $N^\prime$ by $N = \sigma_M - \sigma_1 + 1$, 
and chose the index $n$ to run from $1$ to $N$.
As a consequence of the assumption that $M \gg 1$, the sum in Eq.~(\ref{eq:Z}) can be approximated by an integral, with the result that
\begin{equation}
\label{eq:z-c}
Z \approx \int \, dM e^{\beta (\mu + E_b) M} Z_M.
\end{equation}

To achieve further progress, we make the following assumptions, which are mathematical consequences of the physical ranges of the parameters necessary for the aggregates to form:
\begin{enumerate}
\item Only ``nearest-neighbor'' globule-globule interactions matter. Thus, the globule at $\Xv_m$ only interacts with those at $\Xv_{m\pm1}$.
\item $|\rv_{\sigma_m} - \rv_{\sigma_{m^\prime}}| - 2 a \ll 2 a$ for all such pairs, indicating that the interacting hard-core globules almost touch each other.
\item $|\sigma_n - \sigma_{n-1}| = N/M \gg 1,$ indicating that the bound globules are evenly spaced along the contour of the polymer chain.
\end{enumerate}
The limits of validity of these assumptions, and the corrections that would arise from relaxing them, are explored in the appendix.
As a consequence of Assumption~1, the partition function depends on the coordinates of the globules only through
the combinations $\Xv_m - \Xv_{m - 1}$ for $m = 2 \ldots M$. Furthermore,
using Assumption~2, the inter-globule interaction simplifies via $e^{-\beta V_{m,m-1}} \rightarrow \delta(|\Xv_m - \Xv_{m - 1}| - 2 a)$,
which implements the notion that all pairs of adjacent, mutually impenetrable globules are strongly tethered to one another.
Finally, using Assumption~3, the partition function decomposes into that of
$M$ subchains, with each subchain being composed of $N/M$ segments and each being stretched to a distance $2 a$ (and where $Z_1$ is the partition function for one subchain). Thus, we have:
\begin{widetext}
\begin{equation}
\label{eq:zm-int}
Z_M \approx V \left(4 \pi \right)^{M} (Z_1)^{M}
= V \left(4 \pi \right)^{M} \left( \bigintsss \left[\prod_{n=2}^{N/M} d \rv_n\right] \delta(|\rv_{1} - \rv_{N/M}| - 2 a) e^{- \beta G_{N/M}(\{\rv_n\}) } \right)^{M}.
\end{equation}
\end{widetext}
where $V$ is the volume of the entire system
 and the factor $\left(4 \pi\right)^{M}$ results from the orientational degrees of freedom of the subchains,
\textit{i.e.}, the freedom of the spheres to relocate around one another.

\subsection{A Single Ideal Chain}

For an ideal chain, for which $U = 0$, the free energy $F_{\mathrm{id}} = - k_B T \ln Z_1$ of a single stretched chain 
has the form $F_{\mathrm{id}} = \frac{3}{2} \frac{(2a)^2 k_B T}{(N/M) \ell_0^2}$~\cite{deGennesI},
which from Eq.~(\ref{eq:zm-int}) results in the scaling,
\begin{equation}
\label{eq:zm-id}
Z_{M\mathrm{;id}} \sim e^{- \frac{3}{2} \frac{(2a)^2 M^2}{N \ell_0^2} }.
\end{equation}

To evaluate the mean, variance, and relative width of this (Gaussian) distribution, we
substitute $Z_{M\mathrm{;id}}$ into Eq.~(\ref{eq:z-c}) to obtain the results
\begin{align}
\label{eq:res-id}
\langle M \rangle & = \frac{1}{3} \beta (E_b + \mu) \frac{R_g^2}{(2 a)^2},  \\
\langle (\delta M)^2 \rangle & = \frac{1}{3} \frac{R_g^2}{(2 a)^2}, \\
\label{eq:res-idl}
\frac{\langle (\delta M)^2 \rangle}{\langle M \rangle^2} & = \frac{1}{3} [\beta (E_b + \mu)]^{-2} \frac{(2 a)^2}{R_g^2} \sim \left( \frac{R_g}{a \langle M \rangle} \right)^2,
\end{align}
where for an ideal chain $R_g \equiv \sqrt{N} \ell_0$.
Expression~(\ref{eq:res-idl}) indicates that the relative width of the distribution is small: 
$\frac{\langle (\delta M)^2 \rangle}{\langle M \rangle^2} \ll 1$, which follows because $R_g \ll \langle M \rangle a$, \textit{i.e.}, because the bound globules swell the chain.

\subsection{A Single Self-Avoiding Chain}
The entropic elasticity of self-avoiding chains has two regimes: a 
harmonic regime, in which the chain is stretched much \emph{less} than the radius of
gyration; and an anharmonic regime, in which the chain is stretched much \emph{more} than the radius of
gyration.
In the harmonic regime, $(N/M)^{\nu} \ell_0 \gg a$,
\textit{i.e.}, the radius of each ``blob''~\cite{Pincus1976} comprising the stretched chain is much greater
than the size of the globular particle. In this regime, the globules all fit
inside the blobs of the polymer chain and, therefore, the aggregate is only slightly swollen by the globules. 
By contrast, physical hyaluronan-proteoglycan aggregates are significantly larger than isolated chains,
and therefore correspond to the anharmonic regime.

In this anharmonic regime, for which \mbox{$(N/M)^{\nu} \ell_0 \ll a$}
and the chain is highly stretched~\cite{deGennesI}, we have from Eq.~(\ref{eq:zm-int})
\begin{equation}
\label{eq:zm-hs}
Z_{M\mathrm{;sa}} \sim e^{- C \left (\frac{2a M}{R_g}\right)^\delta},
\end{equation}
where the radius of gyration of the globule-free chain is ${R_g \sim N^\nu \ell_0}$ (for self-avoiding chains, $\nu = 3/5$ according to Flory theory), $\delta \equiv (1-\nu)^{-1} = 5/2$~\cite{deGennesI}, and $C$ is a constant.
Note that $Z_{M\mathrm{;sa}}$ becomes $Z_{M\mathrm{;id}}$ if the ideal chain values $\nu = \frac{1}{2}$ and $C = \frac{3}{2}$ are
substituted into Eq.~(\ref{eq:zm-hs}).
Next, the use of the method of steepest descent~\cite{f1} gives the following expressions for the mean, variance, and relative width of the size distribution of aggregates for the case of self-avoiding polymer chains:
\begin{align}
\label{eq:resm-sa}
\langle M \rangle & \approx \left(\frac{\beta (\mu+E_b)}{C \delta}\right)^{\frac{1}{\delta - 1}} \left( \frac{R_g}{2 a} \right)^{\frac{\delta}{\delta - 1}} \nonumber \\
& = \left(\frac{\beta (\mu+E_b)}{C \delta}\right)^{\frac{1}{\nu} - 1} \left( \frac{R_g}{2 a} \right)^{\frac{1}{\nu}}  \nonumber \\
& = \left(\frac{\beta (\mu+E_b)}{C \delta}\right)^{\frac{1}{\nu} - 1} N \left( \frac{\ell_0}{2 a} \right)^\frac{1}{\nu}
\end{align}
\begin{align}
\label{eq:resv-sa}
\langle (\delta M)^2 \rangle & \approx
\frac{1}{C \delta (\delta - 1) (2 a / R_g)^{\delta} \langle M \rangle^{\delta - 2}} \nonumber \\
& =  \frac{1}{C \delta (\delta - 1)} \left(\beta (\mu+E_b)\right)^{\frac{1}{\nu} - 2} \left( \frac{R_g}{2 a} \right)^{\frac{1}{\nu}}  \nonumber \\
& = \frac{1}{C \delta (\delta - 1)} \left( \beta (\mu+E_b) \right)^{\frac{1}{\nu} - 2} N \left( \frac{\ell_0}{2 a} \right)^\frac{1}{\nu}
\end{align}
\begin{align}
\label{eq:resw-sa}
\frac{\langle (\delta M)^2 \rangle}{\langle M \rangle^2} \sim \left( \beta (\mu+E_b) \right)^{-\frac{1}{\nu}} \left( \frac{R_g}{2 a} \right)^{- \frac{1}{\nu}} \nonumber \\
\sim \langle M \rangle^{-1} \left( \beta (\mu+E_b) \right)^{-1} \ll 1
\end{align}
In particular, if the Flory result $\nu = 3/5$ is inserted, these expressions simplify to Eqs.~(\ref{eq:res}).

\begin{figure}[thp]
\includegraphics[angle=0]{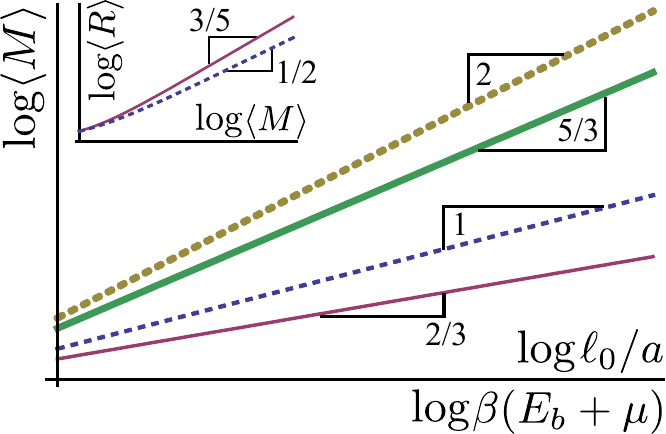}
  \caption{Scaling laws [Eqs.~(\ref{eq:res-id}),~(\ref{eq:resm-sa})] for the number of globules $\langle M \rangle$ in an aggregate 
as a function of $\beta(\mu + E_b)$ (bottom, thin) and $\ell_0/a$ (top, thick), plotted on a log-log scale, with the numbers indicating the line slopes. 
The results from the ideal chain model are dashed, whereas those from the Flory theory of self-avoiding chains are solid.
The inset shows the radius of gyration $\langle R \rangle$ of an aggregate as a function of $\langle M \rangle$:
for sufficiently large aggregates, $\langle R \rangle$ scales as $\langle M \rangle^\nu$. The dashed line
is the result for only nearest-neighbor globule-globule interactions, 
whereas the solid line is the result that includes interactions between all globule pairs.}
\label{fig:scaling}
\end{figure}

\section{Many Aggregates in Suspension}
\label{sec:sol}
To study the structure and thermodynamics of aggregates in suspension,
we first consider the configurations of a single isolated aggregate,
and then include the effects of inter-aggregate interaction.
Thus, let us focus on an aggregate having $M$ bound globules, and consider the configurations 
of the aggregate that dominate $Z_M$.
Under Assumptions 1, 2, and 3, these configurations form a family characterized by the
location of the center of mass of the aggregate, as well as 
the values for the $\sim 2 M$ angular coordinates of $\Xv_m - \Xv_{m-1}$. 
Thus, the configuration space is identical
to that of a freely-jointed polymer chain composed of $M$ segments of segment length $2a$.
As a result, the characteristic linear size of a single aggregate is approximately $2a \sqrt{M}$.

However, Assumption~1 is too strong to describe a physical aggregate. 
Thus, we relax this assumption by allowing for interactions between arbitrary pairs of globules (which are labeled $m$ and $m^\prime$).
There are then two notable effects of these additional interactions.
First, for small $|m - m^\prime|$, the interactions limit the ranges of the bond angles. For example, as a consequence of the fact that two next-nearest-neighbor spheres cannot overlap, the angle between globules $m$, $m+1$, and $m+2$ must lie within the range $\pi/3$ to $5\pi/3$. 
Such effects quickly diminish as $|m - m^\prime|$ increases.
Macroscopically, these interactions have the effect of stiffening the chain at fixed contour length. 
We take this effect into account by representing the aggregates as a freely-jointed chains having a segment length on the order of, but somewhat greater than, $2a$.
Second, the avoidance between distant globules swells the aggregate,
which, therefore, we represent as a self-avoiding chain.
This effect is analogous to that of the swelling of a polymer chain
in good solvent, with the consequent result for the aggregate size: $\langle R \, \rangle \sim 2 a M^\nu$~\cite{Flory, deGennesII}.
The size distribution of the aggregate, given by Eqs.~(\ref{eq:res-id}, \ref{eq:resm-sa}-\ref{eq:resw-sa}),
is not affected by these interactions. 
This is a consquence of the fact that the number $2M$ of orientational degrees of freedom of the effective aggregate chain is small
compared to the total number $3N$ of chain degrees of freedom.

\begin{figure}[thp]
\includegraphics[angle=0]{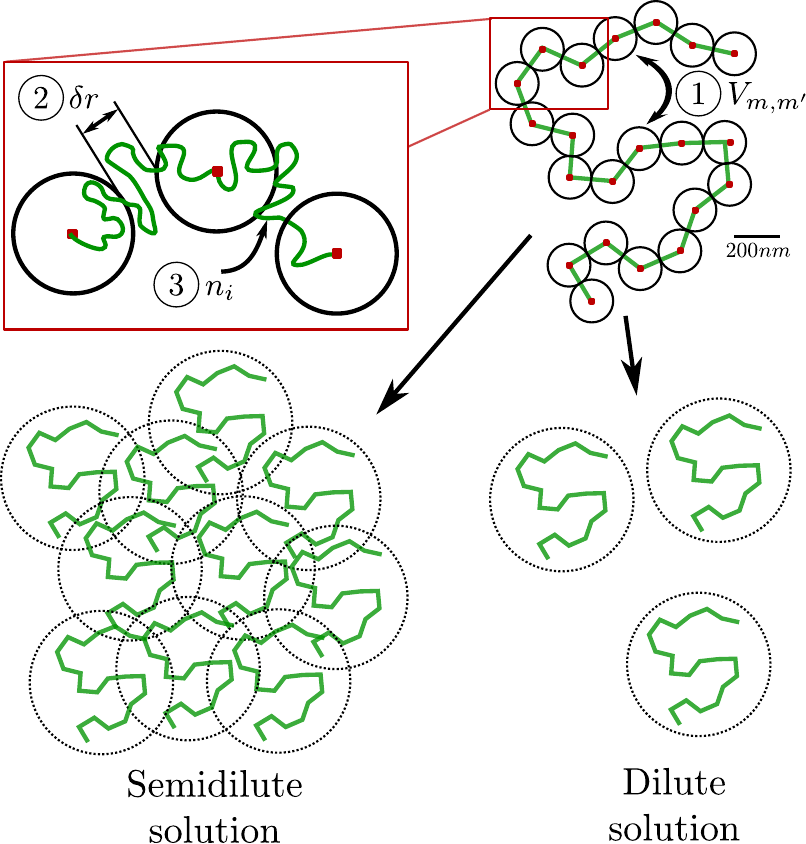}
  \caption{The structure of aggregates in suspension. The aggregates behave as self-avoiding polymer chains
of segment length $2a$, composed of $M$ segments. In the dilute regime, the aggregates do not overlap, as shown on the bottom right. 
In the semi-dilute regime, as shown on the bottom left, the aggregates overlap at small globule packing fractions.
In the dense regime (not shown), the globule packing fraction approaches $1$.
The Assumptions are indicated using the circled numbers $1$, $2$, and $3$.}
\label{fig:sols}
\end{figure}

We now consider many aggregates in suspension, and explore their thermodynamics in three regimes---dilute,
semidilute, and dense.
As we shall see:  (1)~For sufficiently \emph{dilute} suspensions, the properties of the individual
aggregate should be largely independent of the aggregate concentration,
with only small corrections due to aggregate-aggregate interactions.
This results from a separation of energy scales, as the entropy of stretching
is much larger than the aggregate-concentration-dependent free energy. 
On the other hand, (2)~for sufficiently \emph{dense} suspensions,
the suspension behaves essentially as a dense suspension of globules,
with only small corrections due to the presence of the polymer chains.
This results from the fact that the concentration-dependent
free energy is much larger than the entropy of chain stretching.
(3)~The crossover between these two regimes occurs
at densities sufficiently large that the suspension
is described as dense rather than dilute or semidilute.

These results follow by considering a suspension of aggregates, 
and introducing the additional thermodynamic quantities $\rho$ (the aggregate number density), $V$ (the system volume), and $P$ (the number of aggregates).
For convenience, we choose to work at fixed $\rho \equiv P/V$.
We remark that the suspension is considered dilute if
the characteristic size of the aggregate
is much smaller than the inter-aggregate spacing, \textit{i.e.},
if $(M^\nu a) \ll (V/P)^{1/3} = \rho^{-1/3}$; see Fig.~\ref{fig:sols}.
For a dilute aggregate suspension, we may estimate the osmotic pressure $\Pi$ using the virial expansion as
\begin{equation}
\label{eq:pi1}
\Pi = \rho T \left( 1 + \frac{B_2}{v_0} \rho v_0 + \cdots \right),
\end{equation}
where $v_0 \equiv \frac{4}{3} \pi (M^\nu a)^3$ is the volume of a single aggregate,
$\rho v_0 \ll 1$ and $B_2/v_0$ is a numerical virial coefficient
(which for a hard-sphere fluid has the value $4$).
Note that the first term on the right-hand side of Eq.~(\ref{eq:pi1}) 
is independent of aggregate size; a weak dependence on aggregate size results from the (small) second term.
Thus, in the dilute regime, the total Helmholtz free-energy density of the suspension $f$ is given by
\begin{equation}
f = \Pi - \rho T \ln Z_M,
\end{equation}
which is dominated by the large chain-stretching term $- \ln Z_M \sim - (a M/R_g)^{\delta} \gg 1$.
The second term in the virial expansion provides only a small correction to the result, shown in Eq.~(\ref{eq:resm-sa}), 
for $\langle M \rangle$, associated with deswelling of each aggregate.

We now turn to the semidilute regime, in which the suspended aggregates overlap, although
the spacing between individual globules is large, relative to their size; see Fig.~\ref{fig:sols}.
This occurs for the density range $v_0^{-1} \ll \rho \ll a^{-3}$.
By analogy with the ideal chains, a mean-field approximation for the properties of semi-dilute suspensions yields an expression for the 
free-energy density given by~\cite{deGennesIII}
\begin{equation}
\beta f \sim \frac{1}{2} a^{3} (M \rho)^2 + \rho C (2 a M/ R_g)^2.
\end{equation}
As $a^{3} \rho \ll 1$, the two terms are only comparable when $(2 a/R_g)^2 \ll 1$,
which may exist for some range of parameters. However, for the physical
parameters of HA-PG aggregates, $a/R_g$ is of order one.
This means that for semi-dilute suspensions of HA-PG aggregates,
the concentration of aggregates will not affect aggregate size.

The scaling laws resulting from this mean-field theory can be corrected to account for aggregate-density fluctuations in suspensions of self-avoiding aggregates.
To do so, we introduce the stretching exponent $\delta$ ($ = 5/2$); then the expression for the free-energy density becomes (see Ref.~\cite{deGennesIII})
\begin{equation}
\beta f \sim \frac{1}{2} a^{-3} (a^3 M \rho)^{9/4} + \rho C (2 a M/ R_g)^{5/2}.
\end{equation}
These fluctuation corrections suggest that only for parameter values such that
$(a/R_g)^{5/2} M^{1/4} \gg (a^3 \rho)^{5/4}$ would the 
aggregate concentration not affect aggregate size.

For solutions dense in globules, the osmotic pressure of the hard-sphere globules
can dominate over the chain-stretching entropy. 
In that case, the chain degrees of freedom provide small corrections
to a free energy that is essentially the same as the free energy of hard spheres
of radius $a$. That is, sufficiently dense suspensions behave as dense suspensions of globules,
with only small corrections arising from the chain binding.

\section{Discussion and Conclusions}
\label{sec:conc}
\as{We conclude by first discussing the implications of the model developed in this work to hyaluronan-proteoglycan aggregates, and then discussing the implications of this model to other materials, including synthetic ones. For hyaluronan-proteoglycan aggregates, we have obtained the dependence of (i)~the number of bound proteoglycans per aggregate and (ii)~the size of the aggregate on model parameters such as the temperature, binding energy, size, and chemical potential of the globules, as well as the contour and persistence lengths of the hyaluronan chain. These predictions can, in principle, be experimentally tested in both biological and synthetic systems. To clarify these potential tests, we revisit an important assumption of the model, viz.~the assumption that the polymer chain is flexible. As mentioned above, this assumption holds as long as the polymer chain segment between two neighboring bound globules is much longer than the polymer persistence length. This, therefore, sets limits on the number of bound globules in an aggregate for which the model remains applicable. We conjecture that for aggregates outside this limit, the binding of additional globules is suppressed due to the bending stiffness of the polymer chain. To test the scaling laws predicted by the model considered here within a single experiment, one may vary the chemical potential of the globules or, equivalently, the concentration of free globules in solution. Other parameters, such as the length-distribution of synthetic hyaluronan and the size of the aggrecan globules may also be varied from experiment to experiment. This model also leads us to obtain expressions for the free energy of a suspension of many aggregates for several regimes, such as dilute (with respect to the aggregate concentration) or semi-dilute (and still dilute with respect to the globules). We thus predict---in terms of the aggregate parameters---the osmotic pressure and compressibility of the suspension and, in the semi-dilute regime, the presence of a correlation length (observable via scattering), as well as the $9/4$ scaling of the osmotic pressure with the aggregate density. These predictions, too, can be tested in experiments on either synthetic or biological hyaluronan-proteoglycan systems, and can be used to deduce materials parameters of these suspensions. Thus, the thermodynamic results obtained here may lead to advances in the understanding of neuroplasticity as well as the relationship between joint mechanics and materials properties of cartilage.}

The beads-on-a-string model, which we have presented here, is also sufficiently general to describe materials
beyond the example of hyaluronan-proteoglycan aggregates on which we have focused. 
This model combines two basic models of statistical mechanics: the hard-sphere model, which
describes suspensions, and the random walk model, which describes polymer chains.
It thus has the potential to describe any material composed of a suspension of 
large globular particles that have specific binding sites through which the globules bind to flexible polymer chains. 
A possible candidate for such a material, designed to
realize the theory presented here, may be envisioned as a synthetic system composed of
colloidal particles and polymer chains.
For this system to aggregate according to the model we have considered,
the colloidal particles must be functionalized with a single
binding site that has a strong affinity to be bound to a polymer chain,
which could be accomplished, \textit{e.g.}, with complementary strands of DNA~\cite{Mirkin1996}.
The chains must be sufficiently long to accommodate many bound colloids,
and the binding energy must be sufficiently strong to promote aggregate formation,
but not so strong as to lead to irreversible binding.
Within this artificial system, it may be possible to control such parameters as the binding energy (via the DNA strand length and sequence), the temperature, and the colloid concentration and size -- either during the preparation phase of the experiment or during the experiment itself.
A suspension composed of these microscopic aggregates would be a biologically inspired fluid metamaterial, 
and have the remarkable property that the size of the aggregates is reversibly controllable.
Therefore, the model presented in this work could guide the creation of suspensions that have finely tunable viscoelasticity as well as exhibit reversible sedimentation.
By studying the properties of macroscopic suspensions of these colloid-polymer aggregates,
one may gain further understanding of the materials properties that biology has realized as a result of natural selection,
as well as use this understanding for materials engineering.

\section{Acknowledgments}
\label{sec:ack}
This work was supported by NSF grants DMR 12 07026 (AS, PMG) and DMR CAREER 0955811 (JEC), as well as the Georgia Institute of Technology (AS).

\appendix
\section{Revisiting Initial Assumptions}
\label{sec:assum}
We revisit the derivation of Eqs.~(\ref{eq:res-id},\ref{eq:resm-sa}-\ref{eq:resw-sa}), justifying Assumptions~2 and 3 
within a mathematical framework and exploring the corrections to the aggregate partition function that result
from relaxing these assumptions.

First, let us examine a generic pair-interaction potential between globules $V(r)$
and consider the distance between two ``nearest-neighbor'' globules $r \equiv |\Xv_m - \Xv_{m-1}|$ (see Fig.~\ref{fig:sols})
and the statistical distribution of that distance, \textit{i.e.}, $e^{- (\beta V(r) + F(r)) M}$,
where $F(r)$ is the free energy of chain stretching.
This distribution is maximized near $V^\prime(r) = - F^\prime(r)$,
and is sharply peaked for sufficiently large $M$. 
If we consider a sequence of potentials that approach $V_{hc}$,
the solution to this equation must approach $r = 2a$,
showing that indeed the distribution is dominated by small inter-globule separations.

For the hard-sphere potential, the validity of Assumption~2 may be tested directly by calculating $\left\langle r - 2 a \right\rangle$
and higher moments of the distributions $Z_{M\mathrm{;id}}$ or $Z_{M\mathrm{;sa}}$. By definition,
\begin{equation}
\langle r^A \rangle = K^{-1} \int_{2a}^\infty e^{- C \left (\frac{r M}{R_g} \right)^\delta} r^A dr,
\end{equation}
where $K \equiv \int_{2a}^\infty e^{- C \left (\frac{r M}{R_g}\right)^\delta} dr$ is the normalization.
The ideal-chain case is obtained by taking $C = 3/2$, $\delta = 2$, in which case the moments of $r$ may be evaluated exactly, in terms of the Error function. For arbitrary values of $\delta$ in the range $\delta > 1$, the above expectation value can be evaluated using the relation
\begin{align}
\label{eq:int}
\int_{\frac{2aM}{R_g}}^\infty e^{- C x^\delta} x^A dx = &
\frac{1}{\delta C} \left( \frac{2 a M}{R_g} \right)^{A + 1 - \delta} e^{- C \left( \frac{2 a M}{R_g} \right)^\delta} \\
& + \frac{A + 1 - \delta}{\delta C}  \int_{\frac{2aM}{R_g}}^\infty e^{- C x^\delta} x^{A - \delta} dx, 
\nonumber
\end{align}
which is obtained using integration by parts. On the right-hand side of this expression, the first
term is much larger than the second term, and the expression can therefore be used to obtain an asymptotic
expansion for the integral on the left-hand side.
Iterating this relation to next order, we obtain (using $A = 0$ and $1$),
\begin{equation}
\frac{\langle r - 2a \rangle}{2 a} \approx \frac{1}{\delta C} \left(\frac{R_g}{2 a M}\right)^\delta \ll 1.
\end{equation}
Using an additional iteration of Eq.~(\ref{eq:int}) ($A = 0, 1$, and $2$), we obtain
\begin{equation}
\frac{\langle (r - \langle r \rangle)^2 \rangle}{\langle r \rangle^2} 
\approx \frac{1}{\delta^2 C^2} \left(\frac{R_g}{2 a M}\right)^{2 \delta} 
\approx \left( \frac{\langle r - 2 a \rangle}{2a}\right)^2 \ll 1.
\nonumber
\end{equation}
Thus, Assumption~2 is justified on the basis that the deviations of $r$ from the value $2 a$ are small,
both in terms of the expectation value of $r$, and fluctuations around this expectation value.

Finally, let us now relax Assumption~3 by considering the distribution of $n_i \equiv \sigma_n - \sigma_{n-1}$ (see Fig.~\ref{fig:sols}).
If each $n_i$ is considered as an identically distributed independent random variable,
it follows from the central limit theorem that if the mean of the sum of $n_i$ is N,
the mean of each $n_i$ is indeed $N/M$.
To calculate the width of the distribution,
we consider the ensemble of fixed $M$ and varying chain contour length $N$, 
determined by a conjugate variable $p_N$.
Thus, the corresponding partition function is given by the sum over all possible arrangements of globules along the chain:
\begin{equation}
Z_{MN\mathrm{;sa}} \approx \sum_{\{n_i\}_1^M\\n_i = 1}^{N} \delta_{N,\sum n_i} e^{- \sum_{n_i} C \left(\frac{2a}{\ell_0^2 n_i^\nu}\right)^\delta}.
\end{equation}
In the $p_N$ ensemble, after summing over the chain contour lengths $N$, this partition function has the form
\begin{equation}
Z_{Mp_N\mathrm{;sa}} \approx \sum_{\{n_i\}\\n_i = 1}^{N}  e^{- \sum_{n_i} C \left(\frac{2a}{\ell_0^2 n_i^\nu}\right)^\delta} e^{- \beta p_N \sum_{n_i} n_i}.
\end{equation}
For large $n_i$, each sum may be approximated by an integral, giving
\begin{equation}
Z_{Mp_N\mathrm{;sa}} \approx \left(\int_{-\infty}^\infty dn_i e^{- C \left(\frac{2a}{\ell_0^2 n_i^\nu}\right)^\delta - \beta p_N n_i}\right)^M.
\end{equation}
Substituting $\langle n_i \rangle \approx N/M$ for all $i$, we find the ``equation of state''  to be
\begin{equation}
\beta p_N \sim \left(\frac{2 a M}{\ell_0 N}\right)^{\delta}.
\end{equation}
By using this approach, we may also calculate the relative width of the distribution of $n_i$, obtaining,
\begin{equation}
\frac{\langle (n_i - N/M)^2\rangle}{(N/M)^2} \sim \left(\frac{\ell_0 N^\nu}{a M^\nu}\right)^{\delta}  \ll 1,
\end{equation}
which reveals that, indeed, $n_i$ is narrowly distributed around the mean $N/M$.
This narrowness holds for
self-avoiding chains ($\nu = 3/5$) as well as for ideal chains ($\nu = 1/2$), in both cases
 under the condition $\frac{\ell_0 N^\nu}{a M^\nu} \ll 1$, \textit{i.e.},
in the physically relevant regime for the aggregates.


\begin{thebibliography}{17}
\bibitem{f1}
{The steepest descent calculation relies on the approximation
$
\int d x f(x) e^{g(x)} \approx \sqrt{\frac{2 \pi}{- g''(x_0)}} f(x_0),
$
where $g'(x_0) = 0$ is a unique local maximum at the value $x_0$.
This may be extended to calculate the second moment using
$
\left(\int d x e^{g(x)}\right)^{-1} \int d x (x - x_0)^2 e^{g(x)} \approx \frac{1}{- g''(x_0)}.
$}
\end{thebibliography}
\end{document}